\documentstyle[12pt,aaspp4,flushrt,epsfig]{article}

\begin{document}

\title{Measurement of Relativistic Orbital Decay in the \\
  PSR B1534+12 Binary System}
\author{I. H. Stairs\altaffilmark{1}, Z. Arzoumanian\altaffilmark{2},
F. Camilo\altaffilmark{3}, A. G. Lyne\altaffilmark{3}, \\
D. J. Nice\altaffilmark{1}, J. H. Taylor\altaffilmark{1}, 
S. E. Thorsett\altaffilmark{1}, A. Wolszczan\altaffilmark{4}}
\altaffiltext{1}{Joseph Henry Laboratories and Physics Department,
       Princeton University, Princeton, NJ 08544;
ingrid\-@pulsar.princeton.edu, david@pulsar.princeton.edu,
joe@\-pulsar.\-princeton.\-edu, steve@\-pulsar.\-princeton.\-edu}
\altaffiltext{2}{Astronomy Department, Cornell University, Ithaca, NY
14853; arzouman@\-spacenet.\-tn.\-cornell.\-edu}
\altaffiltext{3}{The University of Manchester, NRAL,
Jodrell Bank, Macclesfield, Cheshire SK11 9DL, UK;
  fc@jb.\-man.\-ac.\-uk, agl@jb.man.ac.uk}
\altaffiltext{4}{Department of Astronomy and Astrophysics, Pennsylvania State
University, University Park, PA 16802; alex@astro.psu.edu}
%\vspace*{0.1in}
%\centerline{Submitted to {\it The Astrophysical Journal}.}

\begin{abstract}
We have made timing observations of binary pulsar PSR~B1534+12 with
radio telescopes at Arecibo, Green Bank, and Jodrell Bank.  By
combining our new observations with data collected up to seven years
earlier, we obtain a significantly improved solution for the
astrometric, spin, and orbital parameters of the system. For the first
time in any binary pulsar system, no fewer than five relativistic or
``post-Keplerian'' orbital parameters are measurable with useful
accuracies in a theory-independent way. We find the orbital period of
the system to be decreasing at a rate close to that expected from
gravitational radiation damping, according to general relativity,
although the precision of this test is limited to about 15\% by the
otherwise poorly known distance to the pulsar.  The remaining
post-Keplerian parameters are all consistent with one another and all but
one of them have fractional accuracies better than 1\%.  By assuming
that general relativity is the correct theory of gravity, at least to
the accuracy demanded by this experiment, we find the masses of the
pulsar and companion star each to be $1.339\pm0.003M_\odot$ and the
system's distance to be $d=1.1\pm0.2$~kpc, marginally larger than
the $d\approx0.7\,$kpc estimated from the dispersion measure. The
increased distance reduces estimates of the projected rate of
coalescence of double neutron-star systems in the universe, a quantity
of considerable interest for experiments with terrestrial
gravitational wave detectors such as LIGO.
\end{abstract}

\keywords{pulsars: individual (PSR B1534+12) --- gravitation ---
  binaries: close --- stars:distances}

\section{Introduction} \label{sec:intro}

Pulsars in double-neutron-star binaries provide the best known
laboratories for experimental tests of gravity in the radiative and
strong field regimes.  Timing analysis of pulsar signals allows
measurement of five Keplerian orbital elements as well as a number of
post-Keplerian (PK) orbital parameters.  The PK parameters can be
analyzed using a theory-independent procedure in which the masses of
the two stars are the only dynamically important {\it a priori}
unknowns (\cite{dt92}).  Each of the PK parameters depends on the
masses in a different way; consequently, if any two of them are
measured, the relevant parameters of the two-body system are fully
determined within any gravitational theory.  If three or more PK
parameters can be measured, the overdetermined system can be used to
test the gravitational theory itself.  Measured values of the PK
parameters $\dot\omega$ (rate of advance of periastron), $\gamma$
(time dilation and gravitational redshift) and $\dot{P_b}$ (orbital
period derivative) for binary pulsar PSR~B1913+16 have been found to
be in excellent agreement with the predictions of general relativity
(\cite{tw89,dt91,tay94}).  In particular, the observed value of
$\dot{P_b}$ confirms that the system is losing energy in the form of
quadrupolar gravitational waves at the rate expected according to
general relativity (GR).

An opportunity to repeat and extend this test of relativistic
gravitation was provided by discovery of a second suitable binary
pulsar, PSR~B1534+12, in a 10.1~hour orbit (\cite{wol91a}).
PSR~B1534+12 is significantly brighter than PSR~B1913+16, and its
pulse has a narrow peak, allowing more precise timing measurements.
Because the orbit is nearly edge-on as viewed from the Earth, the PK
parameters $r$ and $s$ (the ``range'' and ``shape'' of the Shapiro
time delay) are more easily measured.  In addition, both $\dot\omega$
and $\gamma$ are measurable, as for PSR B1913+16, because of their
contribution to secularly accumulating effects in the data.  The
resulting overdetermination of the orbit has already led to a
non-radiative test of gravitation theory in the strong-field regime,
complementing the $\dot\omega$-$\gamma$-$\dot P_b$ test for
PSR~B1913+16 (\cite{twdw92}).

Previously published timing measurements of PSR~B1534+12 were all made
with the Arecibo 305\,m telescope.  We have extended this sequence of
observations (\cite{arz95}) through early 1994, when the telescope
went out of normal service for a major upgrading; since then we have
used the 43\,m telescope at Green Bank, West Virginia, and the 76\,m
Lovell telescope at Jodrell Bank, England, to acquire additional data.
Owing to its slower and less eccentric orbit, the expected rate of
orbital period decay for PSR B1534+12 is more than an order of
magnitude less than for PSR B1913+16.  However the timing data are
considerably more precise, and we find that fitting our full sequence
of pulse times of arrival (TOAs) to the standard relativistic pulsar
timing model now yields a measurement of $\dot P_b$ in the
solar-system barycentric rest frame with uncertainty approximately 7\%
of the expected GR value.  The result is a convincing second test of
the dissipative coupling between accelerating masses and a
gravitational radiation field that carries energy and angular momentum
away from the system.  Like the first such test (\cite{tw89,tay94})
these results are in good accord with general relativity theory.  The
precision of the new result is limited by uncertainty in the necessary
kinematic corrections, which depend on the poorly known pulsar
distance.  Interestingly, we can invert the test, assuming that GR is
the correct theory of gravity, and calculate that the distance of
PSR~B1534+12 must be about 1.6 times that estimated from the
dispersion measure (\cite{tc93}).

\section{Observations} \label{sec:obs}
Observations at the 305\,m Arecibo telescope were made primarily with
the Princeton Mark~III system (\cite{skn+92}), using dual-polarization
$2\times32$-channel filterbanks with 0.25\,MHz channels at 430\,MHz
and 1.25\,MHz channels at 1400\,MHz.  The signals in each channel were
square-law detected, and the two polarizations summed in hardware and
then folded at the topocentric pulsar period.  Integration times were
typically 3 minutes at 430\,MHz and 5 minutes at 1400\,MHz.  The
resulting total-intensity profiles have 512 (430\,MHz) or 1024
(1400\,MHz) bins across the 37.9\,ms period.  The effective time
resolution is dominated by the dispersion smearing in a single
channel, approximately 304\,$\mu$s at 430\,MHz and 44\,$\mu$s at
1400\,MHz; quadrature addition of the post-detection time constants
and the boxcar-averaging bin sizes used for these observations yields
the total resolutions quoted in Table~\ref{tab:obsparms}.  The Arecibo
observations extend from August 1990 through March 1994.

We observed PSR B1534+12 with the National Radio Astronomy Observatory
43\,m telescope at Green Bank West Virginia, between March 1994 and
May 1997.  Observing sessions were spaced typically at two month
intervals, and we used frequencies of 575 and 800\,MHz.  A digital
Fourier transform spectrometer (the ``Spectral Processor'') analyzed
signals in each of two polarizations into 512 spectral channels across
a 40\,MHz passband.  The spectra were folded synchronously at the
predicted topocentric pulse period and averaged for three minutes.
The resulting pulse profiles had 128 phase bins, or a resolution of
$296\,\mu$s.  The data were de-dispersed after detection and opposite
polarizations summed to produce a single total-intensity pulse profile
for each integration.

We observed PSR~B1534+12 with the 76\,m Lovell Telescope between
January and July 1997, using the Princeton Mark~IV data acquisition
system at 610\,MHz (\cite{stt98}).  Exploratory observations were made
in January and February, followed by a concentrated campaign from 16
June to 14 July during which time the pulsar was observed for up to 3
hours, during scintillation maxima, nearly every day.  For each sense
of circular polarization, a 5\,MHz bandpass was mixed to baseband
using local oscillators in phase quadrature.  The four resulting
signals were low-pass filtered at 2.35\,MHz (filters 60~dB down at
2.5\,MHz), sampled at 5\,MHz, quantized to 4 bits, and written to a
large disk array and later to magnetic tapes.  Upon playback, the
undetected signal voltages were dedispersed using the phase-coherent
technique described by Hankins and Rickett (1975). \nocite{hr75} After
amplitude calibrations are applied, self- and cross-products of the
right- and left-handed complex voltages yield the Stokes parameters of
the incoming signal.  These products were folded at the topocentric
pulsar period using 2048 phase bins, and pulse TOAs were determined
from the resulting total-intensity profiles using 190\,s integrations.
A summary of the more important parameters and statistics of all four
observing systems is presented in Table~\ref{tab:obsparms}.

We used the same TOA-fitting procedure for all data sets.  Each
observed profile was fitted to a standard template, using a
least-squares method in the Fourier transform domain to measure its
time offset (\cite{tay92}).  The offset was added to the time of the
first sample of a period near the middle of the integration, thereby
yielding an effective pulse arrival time.  A different standard
template was used for each observing system and frequency; they were
made by averaging the available profiles over several hours or more.
Uncertainties in the TOAs were estimated from the least squares
procedure, and also from the observed scatter of the TOAs within 30
minutes of each one.  Each observatory's local time standard was
corrected retroactively to the UTC timescale, using data from the
Global Positioning System (GPS) satellites.

\section{Data Analysis}
\subsection{The Timing Model}\label{sec:model}

A pulse received on Earth at topocentric time $t$ is emitted at a time
in the comoving pulsar frame given by
\begin{equation}
T =  t-t_0+\Delta_C-D/f^2 + \Delta_{R\odot} + \Delta_{E\odot}
  -\Delta_{S\odot} - \Delta_R - \Delta_E - \Delta_S\,.
\label{eqn:orbit}
\end{equation}
Here $t_0$ is a reference epoch and $\Delta_C$ is the offset between
the observatory master clock and the reference standard of terrestrial
time.  The dispersive delay is $D/f^2$, where $D={\rm
DM}/2.41\times10^{-4}$, with dispersion measure DM in cm$^{-3}$pc,
radio frequency $f$ in MHz, and the delay in seconds.  Finally,
$\Delta_{R\odot}$, $\Delta_{E\odot}$, and $\Delta_{S\odot}$ are
propagation delays and relativistic time adjustments for effects
within the solar system, and $\Delta_R$, $\Delta_E$ and $\Delta_S$ are
similar terms accounting for phenomena within the pulsar's orbit
(\cite{dd86,tw89,dt92}).  (We ignore the uninteresting constant and
uniform rate of change of the overall propagation delay.)  The orbital
$\Delta$ terms are defined by:
\begin{eqnarray}
\Delta_R & = & x \sin\omega (\cos u -e)  + x ( 1-e^2)^{1/2}
       \cos\omega \sin u, \\
\Delta_E & = & \gamma \sin u, \\
\Delta_S & = & -2r \ln \left\{ 1-e\cos u - s \left[
  \sin\omega (\cos u - e) + (1-e^2)^{1/2} \cos\omega \sin u
  \right] \right\}\,. 
\end{eqnarray}
These are written in terms of the
eccentric anomaly $u$ and true anomaly $A_e(u)$, and the time
dependence of $\omega$, which are related by:
\begin{eqnarray}
u-e\sin u & = & 2\pi \left[ \left( {{T-T_0}\over{P_b}} \right) -
  {{\dot P_b}\over 2} \left( {{T-T_0}\over{P_b}} \right)^2
  \right], \\
A_e(u) & = & 2 \arctan \left[ \left( {{1+e}\over{1-e}}
  \right)^{1/2} \tan {u\over2} \right], \\
\omega & = & \omega_0 + \left(\frac{P_b\,\dot{\omega}}{2\pi}\right) 
         A_e(u).  
\end{eqnarray}
At a given time $t$, then, the propagation delay across the pulsar
orbit is calculated by a model which incorporates ten parameters
implicitly defined in the above equations: five Keplerian parameters
($x$, $\omega$, $T_0$, $P_b$, $e$) and five PK parameters
($\dot{\omega}$, $\dot{P_b}$, $\gamma$, $r$, $s$).  These quantities,
in conjunction with a simple time polynomial to model the spin of the
pulsar and with astrometric parameters to model the propagation of the
signal across the solar system, constitute the free parameters to be
fit in the theory-independent timing model.

In a particular theory of gravity, the five PK parameters can be
written as functions of the pulsar and companion star masses, $m_1$
and $m_2$, and the well-determined Keplerian parameters.  Of
particular interest is general relativity, where the equations are as
follows (see \cite{dd86,tw89,dt92}):
\begin{eqnarray}
\dot\omega &=& 3 \left(\frac{P_b}{2\pi}\right)^{-5/3}
  (T_\odot M)^{2/3}\,(1-e^2)^{-1}\,, \label{eq:omdot} \\
\gamma &=& e \left(\frac{P_b}{2\pi}\right)^{1/3}
  T_\odot^{2/3}\,M^{-4/3}\,m_2\,(m_1+2m_2) \,, \\
\dot P_b &=& -\,\frac{192\pi}{5}
  \left(\frac{P_b}{2\pi}\right)^{-5/3}
  \left(1 + \frac{73}{24} e^2 + \frac{37}{96} e^4 \right)
  (1-e^2)^{-7/2}\,T_\odot^{5/3}\, m_1\, m_2\, M^{-1/3}\,,
  \label{eq:pbdot} \\
r &=& T_\odot\, m_2\,, \\
s &=& x \left(\frac{P_b}{2\pi}\right)^{-2/3}
  T_\odot^{-1/3}\,M^{2/3}\,m_2^{-1}\,. \label{eq:s}
\end{eqnarray}
Here the masses $m_1$, $m_2$, and $M\equiv m_1+m_2$ are expressed in
solar units, and we use the additional shorthand notations
$s\equiv\sin i$ and $T_\odot\equiv GM_\odot/c^3 = 4.925490947\,\mu$s,
where $i$ is the angle between the orbital angular momentum and the
line of sight, $G$ the Newtonian constant of gravity, and $c$ the
speed of light.

\subsection{Arrival Time Analysis}

We used the standard {\sc tempo} analysis software (\cite{tw89}; see
also Internet location http:$/\!/$pulsar.princeton.edu/tempo) together
with the JPL~DE200 solar-system ephemeris (\cite{sta90}) to fit the
measured pulse arrival times to the timing model with a least-squares
technique.  Results for the astrometric, spin, and dispersion
parameters of PSR B1534+12 are presented in Table~2.  Initial tests
indicated the likely presence of systematic errors in the Arecibo data
taken at 430\,MHz (see below), so these data were used only for the
calculation of the dispersion correction.  The uncertainty quoted for
DM is dominated by the unavoidable difficulty in aligning the
frequency-dependent pulse shapes obtained in the different observing
bands.  We list for the first time a significant measurement of the
rate of change of dispersion measure for this pulsar, determined from
the dual-frequency data collected at Arecibo from 1990 to 1994.  We
note that there is no guarantee that this parameter will have remained
constant since 1994.

We fit the data to two models of the pulsar orbit.  The
theory-independent ``DD'' model (\cite{dd86}) treats all five PK
parameters defined in \S3.1 as free parameters in the fit.
Alternatively, the ``DDGR'' model (\cite{tay87a,tw89}) assumes general
relativity to be correct and uses equations~\ref{eq:omdot}
through~\ref{eq:s} to link the PK parameters to $M\equiv m_1+m_2$ and
$m_2$; consequently it requires only two post-Keplerian free
parameters.

Table~\ref{tab:orbparms} presents our adopted orbital parameters.
Uncertainties given in the table are approximately twice the formal
``$1\,\sigma$'' errors from the fit; we believe them to be
conservative estimates of the true 68\%-confidence uncertainties,
including both random and systematic effects.  The Keplerian orbital
parameters include the period $P_{b}$, projected semi-major axis
$x\equiv a_1\sin i/c$, eccentricity $e$, longitude of periastron
$\omega$, and time of periastron $T_0$.  These quantities are followed
by the measured post-Keplerian parameters relevant to each of the two
models.  For the first time, the precision of this experiment requires
the DDGR solution to include a parameter we call ``excess $\dot
P_b$,'' which accounts for an otherwise unmodeled acceleration
resulting from galactic kinematics.

The best estimates of the masses of the pulsar and its companion come
from the DDGR solution.  We find the masses to be equal:
$m_1=m_2=1.339\pm0.003~M_\odot$.  For the sake of comparison
Table~\ref{tab:orbparms} lists, in italic numbers, computed PK
parameter values corresponding to the measured masses in the DDGR fit.
We call attention to the fact that the fitted and derived parameter
values are in excellent accord, indicating good agreement of the
theory-independent solution with general relativity.

\begin{figure}[htb]
\centerline{\epsfig{file=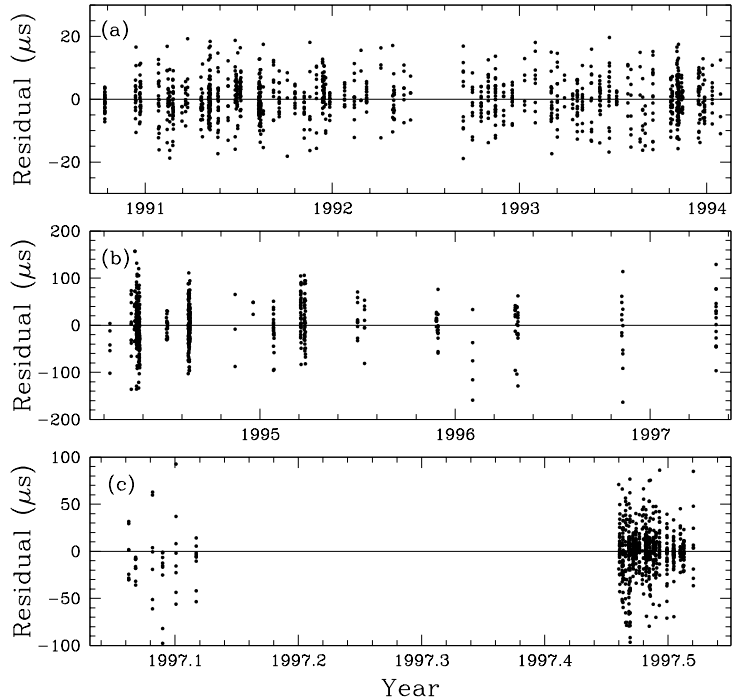,width=4in}}
\figcaption{Post-fit residuals versus date for 
(a) Arecibo 1400\,MHz, (b) Green Bank and (c) Jodrell Bank data.
\label{fig:dayres}}
\end{figure}

\begin{figure}[htb]
\centerline{\epsfig{file=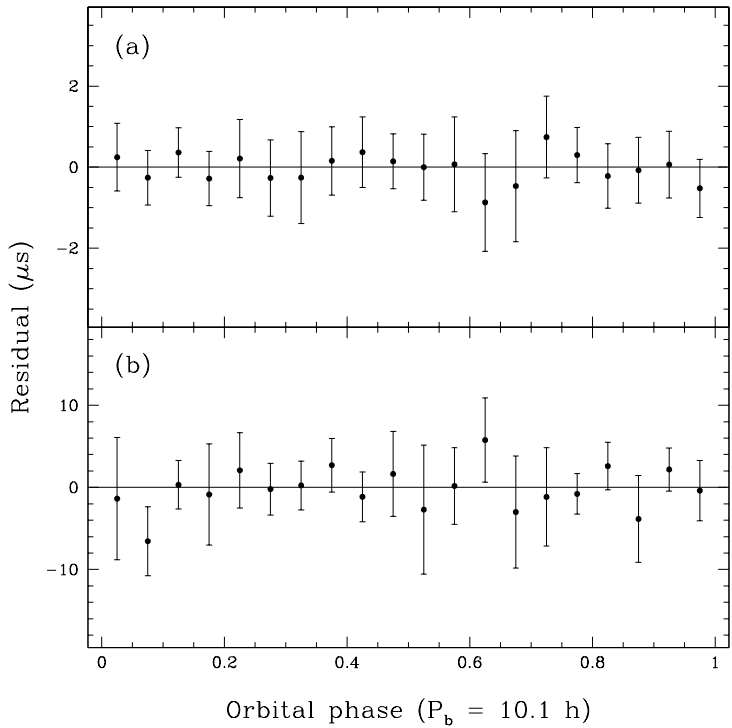,width=4in}}
\figcaption{Average post-fit residuals as a function of
orbital phase for (a) the Arecibo 1400\,MHz data and (b) the Jodrell
Bank 610\,MHz data. \label{fig:orbbin}}
\end{figure}

Figure 1 shows the post-fit residuals for the Arecibo 1400\,MHz data,
the Green Bank data, and the Jodrell Bank data, plotted as functions
of date.  Even within a single data set, the TOA uncertainties can
vary by factors of three or more, above and below the median values
$\sigma_{\rm TOA}$ listed in Table~1, because of scintillation-induced
intensity variations.  We have not attempted to show these differences
in data quality by means of error bars in the residual plots.
Figure~2 illustrates the average post-fit residuals for the Arecibo
measurements at 1400\,MHz and the Jodrell Bank data at 610\,MHz,
plotted as functions of orbital phase.

\subsection{Data Confidence Tests}

For some time we have suspected that solutions for the fitted
parameters of PSR B1534+12, as determined from the Arecibo 430\,MHz
data, are biased by small systematic errors in the TOAs.  These errors
likely arise from imperfect post-detection dispersion removal that
follows as a consequence of the relatively coarse frequency resolution
of the filter-bank spectrometer.  It is extremely difficult to achieve
TOAs with systematic errors less than a few percent of the dominant
instrumental smoothing effects in the data---in this instance, the
dispersion time across a single channel.  Variable spectral features
caused by interstellar scintillation, together with slightly irregular
filter passbands and center frequencies, make the Arecibo 430\,MHz
measurements unreliable at the several microsecond level.

\begin{figure}[htb]
\centerline{\epsfig{file=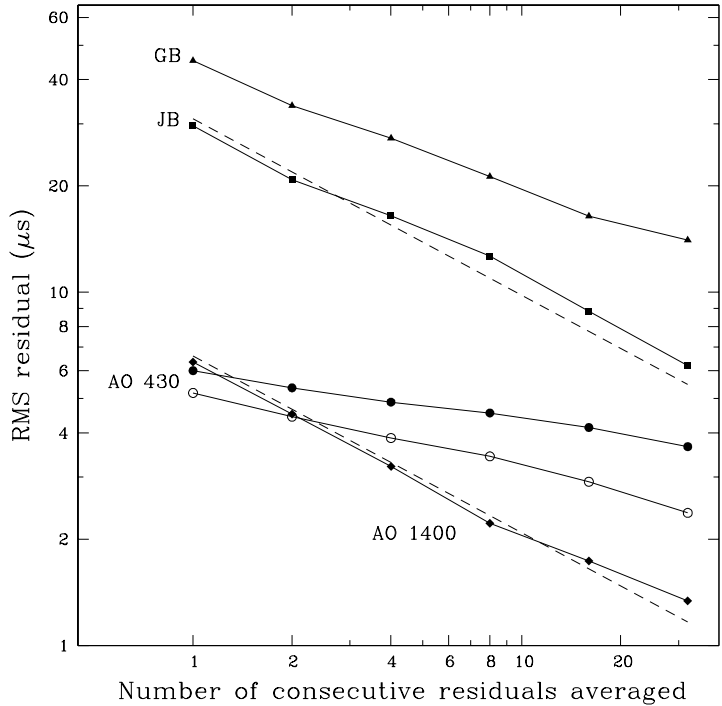,width=4in}}
\figcaption{Root-mean-square residual versus number of
consecutive residuals averaged, for the four data sets: Green Bank,
Jodrell Bank, and Arecibo at 430 and 1400\,MHz.  For the Arecibo
430\,MHz data, filled circles represent the DD solution presented in
Table 3, and open circles represent a DDGR fit to the 430\,MHz data
alone.  Dashed lines indicate the expected slope of $-1/2$ for
uncorrelated residuals. \label{fig:resn}}
\end{figure}

Some further information on the data quality is presented in Figure~3.
Here we follow Taylor and Weisberg (1989) and explore the statistical
properties of the post-fit residuals to see if they ``integrate down''
as $n^{-1/2}$ when $n$ consecutive values are averaged.  As expected,
this test confirms that the Arecibo 430\,MHz data (and to a lesser
extent the Green Bank data, which anyway receive low weight) deviate
from the ideal $n^{-1/2}$ behavior.  This indicates the likely
presence of systematic errors in the TOAs, though we cannot completely
rule out the possibility of a frequency-dependent deviation from the
timing model (e.g., aberration-induced pulse profile variations).  In
any case, it is clear that for the highest accuracy within our
preferred models we must omit the existing Arecibo 430\,MHz
measurements from our solution, as described above.

The parameters of PSR B1534+12 are therefore best determined from the
Arecibo 1400\,MHz data along with the Green Bank and Jodrell Bank data
sets.  The higher observing frequency of 1400\,MHz assures that for
these measurements the total instrumental smoothing, even with
1.25\,MHz channel bandwidths, is only 97\,$\mu$s (see Table~1).
Better still, the Princeton Mark IV system used at Jodrell Bank was
explicitly designed to minimize or eliminate dispersion-related
systematic errors in TOAs.  We used these two data sets, both with and
without the much lower-weighted Green Bank data, to measure and check
the system parameters with highest accuracy.  Arbitrary offsets were
allowed between the different data sets to allow for
frequency-dependent changes in pulse shape and slight differences in
the standard profile alignments.  For obvious reasons, we look forward
to making improved measurements at 430\,MHz with the Princeton Mark IV
system, when the Arecibo telescope is available for use following its
upgrade.

Note that each of the data sets treated in Figure~3 may deviate at
least slightly from the ideal $n^{-1/2}$ behavior, suggesting small
systematic errors of unknown origin.  To partially compensate for this
effect, uncertainties in TOAs were increased (in quadrature) by 2.9,
17.1, and 20.5\,$\mu$s, respectively, when calculating weights for the
Arecibo, Jodrell Bank, and Green Bank data sets.

\section{Discussion} \label{sec:disc}

\subsection{Observed Change in Orbital Period:  Distance to PSR\,B1534+12}

Before proceeding to tests of GR we must correct the observed $\dot
P_b$, which is expressed in the reference frame of the solar system
barycenter, to the center-of-mass frame of the binary pulsar.  Damour
and Taylor (1991) \nocite{dt91} derived an expression for the most
significant bias, which arises from galactic kinematic effects.  It
can be written as the sum of terms arising from acceleration toward
the plane of the Galaxy, acceleration within the plane of the Galaxy,
and an apparent acceleration due to the proper motion of the binary
system:
\begin{equation}\label{eqn:gal}
\left(\frac{\dot{P_b}}{P_b}\right)^{\rm gal} = -\,\frac{a_z\sin b}{c}
 \,-\,\left[\frac{v_0^2}{cR_0} \cos l +\frac{v_1^2}{cR_1}
 \cos\lambda\right] 
 +\mu^2\frac{d}{c}.
\end{equation}
Here $a_z$ is the vertical component of galactic acceleration, $l$ and
$b$ the galactic coordinates of the pulsar, $\mu$ the proper motion,
$v_0$ the circular velocity at the Sun's galactocentric radius $R_0$,
$v_1$ and $R_1$ the corresponding values at the pulsar's location,
$\lambda$ the angle between the Sun and the galactic center as seen
from the pulsar, and $d$ the distance from the pulsar to the Sun.  The
pulsar distance can be estimated from the dispersion measure, together
with a smoothed-out model of the free electron distribution in the
Galaxy.  The Taylor and Cordes (1993) \nocite{tc93} model yields
$d\approx0.7$\,kpc for PSR B1534+12, with an uncertainty of perhaps
0.2\,kpc.  At this distance we estimate
$a_z/c=(1.60\pm0.13)\times10^{-19}\,\mbox{s}^{-1}$ from the model of
Kuijken and Gilmore (1989). \nocite{kg89} Following Damour and Taylor
(1991)\nocite{dt91}, we assume a flat galactic rotation curve and take
$v_0=v_1=222\pm20\,$km\,s$^{-1}$ and $R_0 = 7.7\pm0.7$\,kpc.  Then,
summing the terms in equation~(\ref{eqn:gal}) and multiplying by $P_b$, we
find the total kinematic correction to be
\begin{equation}
\left(\dot P_b\right)^{\rm gal} = (0.038\pm0.012)\times10^{-12}\,.
\end{equation}
The uncertainty in this correction is dominated by the uncertainty in
distance, which is only roughly estimated by the Taylor and Cordes
model.

Our measurement of the intrinsic rate of orbital period decay is
therefore
\begin{equation}
\left(\dot P_b\right)^{\rm obs} - \left(\dot P_b\right)^{\rm gal}
        = (-0.167\pm0.018)\times 10^{-12}\,.
\end{equation}
Under general relativity, the orbital period decay due to
gravitational radiation damping, $(\dot{P_b})^{\rm GR}$, can be
predicted from the masses $m_1$ and $m_2$ (eq.~\ref{eq:pbdot}), which
in turn can be deduced from the high precision measurements of
$\dot{\omega}$ and $\gamma$.  The expected value is
\begin{equation}
\left(\dot P_b\right)^{\rm GR} = -0.192\times 10^{-12}\,.
\end{equation}
Although the measured value in the pulsar center-of-mass frame differs
from this prediction by 1.4 standard deviations, it can be brought
into good agreement by increasing the pulsar distance to slightly over
1\,kpc.  Stated another way, we can assume that GR is the correct
theory of gravity, measure the ``excess $\dot P_b$'' for the system as
described above and presented in Table~3, and then invert
equation~\ref{eqn:gal} to determine the pulsar distance (\cite{bb96}).  By
this method we obtain $d=1.1\pm0.2$~kpc (68\% confidence limit).  The
uncertainty is dominated by the measurement uncertainty of $(\dot
P_b)^{\rm obs}$, rather than uncertainties in the galactic rotation
parameters or the acceleration $a_z$.  It follows that continued
timing of this system should lead to a much more precise distance
measurement, in due time.

We note that the timing solution provides a second, independent
constraint on the distance.  The upper limit on parallax,
$\pi<1.7$\,mas (Table 2) constrains the distance to $d>0.6$\,kpc.
While the parallax distance has less precision than the kinematic
distance, it is reassuring that these measurements are in agreement.

An accurate distance for PSR~B1534+12 is of considerable interest
because of the importance of this system to estimates of the rate of
coalescence of binary neutron-star pairs in a typical galaxy. Previous
estimates of this rate have used much smaller distances for
PSR~B1534+12, including 0.4~kpc (\cite{nps91}), 0.5~kpc
(\cite{phi91}), and 0.7~kpc (\cite{cl95,vl96}), leading to a low
estimate of its intrinsic luminosity.  If our newly determined
distance is correct, as we believe it must be, the earlier studies
have overestimated the number density of systems like PSR~B1534+12 in
the universe by factors of 2.5 to 20. We find that the most recent
observational estimate of the double neutron star inspiral rate,
$2.7\times10^{-7}\mbox{yr}^{-1}$ (\cite{vl96}), must be reduced to
between $1.1\times10^{-7}\mbox{yr}^{-1}$ and
$1.4\times10^{-7}\mbox{yr}^{-1}$, depending on the unknown scale
height of such systems in the Galaxy.  This is, of course, a lower
bound on the number of coalescence events; the actual rate will be
higher by unknown factors that account for the fraction of pulsars
with beams that do not cross the Earth and the number of double
neutron star binaries that do not contain an active pulsar.

\subsection{Test of Relativity}

\begin{figure}[htb]
\centerline{\epsfig{file=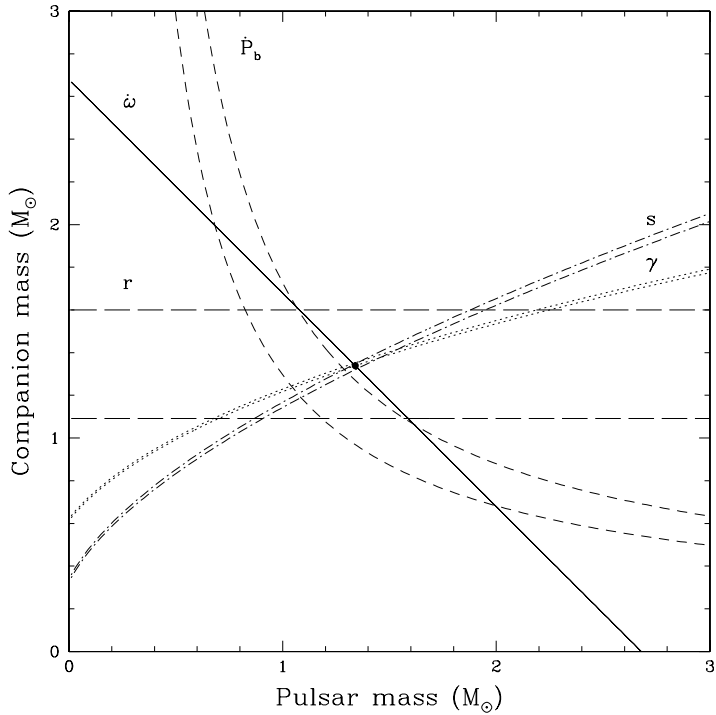,width=4in}}
\figcaption{Mass-mass diagram for the PSR B1534+12
system.  Labeled curves illustrate 68\% confidence ranges of the DD
parameters listed in Table 3.  The filled circle denotes the component
masses according to the DDGR solution.  A kinematic correction for
assumed distance $d=0.7\pm0.2\,$kpc has been subtracted from the
observed value of $\dot{P_b}$.  A slightly larger distance removes the
small apparent discrepancy. \label{fig:massmass}}
\end{figure}

Our analysis of this experiment represents the second test of general
relativity based on the $\dot\omega$, $\gamma$, and $\dot P_b$
parameters of a binary pulsar system, and the first one to add
significant measurements of the Shapiro-delay parameters $r$ and $s$.
The left-hand sides of equations~(\ref{eq:omdot}--\ref{eq:s}) represent
measured quantities, as specified for this experiment in the ``DD''
column of Table~3.  If GR is consistent with the measurements and
there are no significant unmodeled effects, we should expect the five
curves corresponding to equations~(\ref{eq:omdot}--\ref{eq:s}) to
intersect at a single point in the $m_1$-$m_2$ plane.  A graphical
summary of the situation for PSR B1534+12 is presented in
Figure~\ref{fig:massmass}, in which a pair of lines delimit the 68\%
confidence limit for each PK parameter (a single line for
$\dot\omega$, whose uncertainty is too small to show).  A filled
circle at $m_1=m_2=1.339~M_\odot$ marks the DDGR solution of Table~3,
and its location on the $\dot\omega$ line agrees well (better than
1\%) with the measured DD values of $\gamma$ and $s$.  We have already
noted that the DD value of $\dot P_b$ is slightly too small when
corrected to the dispersion-estimated distance.  However, as discussed
above, this discrepancy can be removed by invoking a larger distance
to the pulsar.  Finally, the value of $r$, although presently little
better than a 20\% measurement, is also well centered on its expected
value.  Altogether we find the measured parameters of the PSR B1534+12
system to be in excellent accord with general relativity, and
consequently this theory has passed another very significant
astrophysical test.

\acknowledgements

The Arecibo Observatory, a facility of the National Astronomy and
Ionosphere Center, is operated by Cornell University under a
cooperative agreement with the National Science Foundation.  The
National Radio Astronomy Observatory is operated by Associated
Universities, Inc., under a cooperative agreement with the US National
Science Foundation.  We thank Jon Bell and Christopher Scaffidi for
assistance with observations.  I.~H.~S. gratefully acknowledges the
support of an NSERC 1967 fellowship.  F.~C. is a Marie Curie Fellow.
S.~E.~T. is an Alfred P.~Sloan Foundation Research Fellow.

\clearpage

%\bibliographystyle{apj1c}
%\bibliography{journals1,modrefs,psrrefs}

\begin{thebibliography}{}
 
\bibitem[Arzoumanian 1995]{arz95}
Arzoumanian, Z. 1995.
\newblock PhD thesis, Princeton University
 
\bibitem[Bell \& Bailes 1996]{bb96}
Bell, J.~F. \& Bailes, M. 1996, ApJ, 456, L33
 
\bibitem[Curran \& Lorimer 1995]{cl95}
Curran, S.~J. \& Lorimer, D.~R. 1995, MNRAS, 276, 347
 
\bibitem[Damour \& Deruelle 1986]{dd86}
Damour, T. \& Deruelle, N. 1986, Ann. Inst. H. Poincar\'e (Physique
  Th\'eorique), 44, 263
 
\bibitem[Damour \& Taylor 1991]{dt91}
Damour, T. \& Taylor, J.~H. 1991, ApJ, 366, 501
 
\bibitem[Damour \& Taylor 1992]{dt92}
Damour, T. \& Taylor, J.~H. 1992, Phys. Rev. D, 45, 1840
 
\bibitem[Hankins \& Rickett 1975]{hr75}
Hankins, T.~H. \& Rickett, B.~J. 1975, Meth. Comp. Phys., 14, 55
 
\bibitem[Kuijken \& Gilmore 1989]{kg89}
Kuijken, K. \& Gilmore, G. 1989, MNRAS, 239, 571
 
\bibitem[Narayan, Piran, \& Shemi 1991]{nps91}
Narayan, R., Piran, T., \& Shemi, A. 1991, ApJ, 379, L17
 
\bibitem[Phinney 1991]{phi91}
Phinney, E.~S. 1991, ApJ, 380, L17
 
\bibitem[Stairs, Taylor, \& Thorsett 1998]{stt98}
Stairs, I.~H., Taylor, J.~H., \& Thorsett, S.~E. 1998, ApJ,
\newblock in preparation
 
\bibitem[Standish 1990]{sta90}
Standish, E.~M. 1990, A\&A, 233, 252
 
\bibitem[Stinebring {\it et al.}  1992]{skn+92}
Stinebring, D.~R., Kaspi, V.~M., Nice, D.~J., Ryba, M.~F., Taylor, J.~H.,
  Thorsett, S.~E., \& Hankins, T.~H. 1992, Rev. Sci. Instrum., 63, 3551
 
\bibitem[Taylor 1987]{tay87a}
Taylor, J.~H. 1987, in { General Relativity and Gravitation}, ed.\ M.~A.~H.
  MacCallum, (Cambridge: Cambridge University Press), 209
 
\bibitem[Taylor 1992]{tay92}
Taylor, J.~H. 1992, Philos. Trans. Roy. Soc. London A, 341, 117
 
\bibitem[Taylor 1994]{tay94}
Taylor, J.~H. 1994, Rev. Mod. Phys., 66, 711
 
\bibitem[Taylor \& Cordes 1993]{tc93}
Taylor, J.~H. \& Cordes, J.~M. 1993, ApJ, 411, 674
 
\bibitem[Taylor \& Weisberg 1989]{tw89}
Taylor, J.~H. \& Weisberg, J.~M. 1989, ApJ, 345, 434
 
\bibitem[Taylor {\it et al.}  1992]{twdw92}
Taylor, J.~H., Wolszczan, A., Damour, T., \& Weisberg, J.~M. 1992, Nature, 355,
  132
 
\bibitem[van~den Heuvel \& Lorimer 1996]{vl96}
van~den Heuvel, E. P.~J. \& Lorimer, D.~R. 1996, MNRAS, 283, L37
 
\bibitem[Wolszczan 1991]{wol91a}
Wolszczan, A. 1991, Nature, 350, 688
 
\end{thebibliography}

\clearpage

\begin{table*}
\caption{\label{tab:obsparms} Parameters of the four observing systems.}
\vspace{2mm}

\begin{center}
\renewcommand{\tabcolsep}{1.4mm}
\begin{tabular}{lrrrr}
\tableline
\tableline
& Arecibo & Arecibo & Green Bank & Jodrell Bank \\ 
& Mark III & Mark III & Spec. Proc. & Mark IV \\ 
\tableline 
Frequency (MHz) \dotfill  & 430 & 1400 & 575, 800 & 610 \\
Bandwidth (MHz) \dotfill  & 8 & 40 & 40 & 5 \\
Spectral Channels \dotfill  & 32 & 32 & 512 & 1 \\
Dedispersing system \dotfill & incoherent & incoherent 
  & incoherent & coherent \\
Time resolution ($\mu$s) \dots  & 329 & 97 & 296 & 19 \\
Integration time (s) \dotfill  & 180 & 300 & 180 & 190 \\
Dates \dotfill &1990.7--94.2 &1990.8--94.1 & 1994.2--97.4 & 1997.0--97.6 \\
Number of TOAs \dotfill  & 2311 & 1170 & 685 & 780 \\
Median $\sigma_{\rm TOA}$ ($\mu$s) \dotfill  & 4.9 & 5.4 & 40 & 21 \\
\tableline
\end{tabular}
\end{center}
\end{table*}
 
\begin{table*}
\caption{\label{tab:astspin} Astrometric, spin, and dispersion
parameters for PSR B1534+12\tablenotemark{a}.}
\vspace{2mm}
\begin{center}
\begin{tabular}{ll}
\tableline
\tableline
Right ascension, $\alpha$ (J2000)\dotfill  & 
  $15^{\rm h}\,37^{\rm m}\,09\fs95994(2)$ \\
Declination, $\delta$ (J2000)  \dotfill  & 
  $11^\circ\,55'\,55\farcs6561(3)$ \\
Proper motion in R.A., $\mu_\alpha$ (mas\,yr$^{-1}$)  \dotfill  & 
1.3(3) \\
Proper motion in Dec., $\mu_\delta$ (mas\,yr$^{-1}$)  \dotfill  & 
$-$25.5(4) \\
Parallax, $\pi$ (mas) \dotfill & $<1.7$ \\
\\
Pulse period, $P$ (ms)  \dotfill  & 
37.9044404878552(5)\\
Period derivative, $\dot P$ $(10^{-18})$ \dotfill & 2.42253(3) \\
Epoch (MJD) \dotfill & 48778.0 \\
\\
Dispersion measure, DM (cm$^{-3}$pc) \dotfill   & 
11.619(12) \\
DM derivative (cm$^{-3}\mbox{pc\,yr}^{-1}$)  \dotfill  & 
$-$0.00036(2) \\
\\
Galactic longitude $l$ (deg) \dotfill  & {\it 20.0} \\ 
Galactic latitude $b$ (deg) \dotfill  & {\it 47.8}  \\
Composite proper motion, $\mu$ (mas\,yr$^{-1}$) \dots & {\it 25.5(4)} \\
Galactic position angle of $\mu$ (deg) \dotfill  & {\it 238.7(2)}  \\
\tableline
\end{tabular}
\end{center}
\tablenotetext{a}{Figures in parentheses are uncertainties in the last digits
quoted, and italic numbers represent derived quantities.}
\end{table*}

\begin{table*}
\caption{\label{tab:orbparms}
Orbital parameters of PSR B1534+12 in the DD and DDGR models\tablenotemark{a}.}
\vspace{2mm}
\begin{center}
\begin{tabular}{lll}
\tableline
\tableline
& DD model & DDGR model \\
\tableline
Orbital period, $P_b$ (d) \dotfill  & 0.42073729929(4) & 0.42073729930(4) \\ 
Projected semi-major axis, $x$ (s) \dotfill  & 3.729463(3) & 3.7294628(7) \\
Eccentricity, $e$ \dotfill  & 0.2736777(5) & 0.2736776(2) \\ 
Longitude of periastron, $\omega$ (deg) \dotfill  & 267.44746(16) 
  & 267.44760(12) \\
Epoch of periastron, $T_0$ (MJD) \dotfill  & 48777.82595097(6) 
  & 48777.82595096(6) \\
\\
Advance of periastron, $\dot\omega$ (deg\,yr$^{-1}$) \dotfill  
  & 1.75576(4) & {\it 1.75577} \\
Gravitational redshift, $\gamma$ (ms) \dotfill  & 2.066(10) & {\it 2.057} \\
Orbital period derivative, $(\dot P_b)^{\rm obs}$ $(10^{-12})$ \dots  &
  $-$0.129(14) & {\it $-$0.1924} \\
Shape of Shapiro delay, $s$ \dotfill  & 0.982(7) & {\it 0.9797} \\
Range of Shapiro delay, $r$ ($\mu{\rm s}$) \dotfill  & 6.7(1.3) & 
 {\it 6.60} \\
Derivative of $x$, $\left |\dot x \right |$ $(10^{-12})$ \dotfill & $<2.5$
& $<0.022$ \\
Derivative of $e$, $\left |\dot e \right |$ $(10^{-15}\,{\rm s}^{-1})$
\dotfill & $<6$ & $<6$ \\
\\
Total mass, $M=m_1+m_2$ ($M_{\odot}$) \dotfill  & \dots & 2.67838(8) \\
Companion mass, $m_2$ ($M_{\odot}$) \dotfill    & \dots & 1.339(3) \\
Excess $\dot P_b$ $(10^{-12})$ \dotfill         & \dots & 0.062(14) \\
\tableline
\end{tabular}
\end{center}
\tablenotetext{a}{Figures in parentheses are uncertainties in the last digits
quoted.  Italic numbers represent derived parameters, assuming
general relativity.}
\end{table*}

\end{document}